\documentstyle[12pt,epsfig]{article}
\newlength{\dinwidth}
\newlength{\dinmargin}
\newcommand{\ra}{\rightarrow}

\newcommand{\ba}{\begin{array}}
\newcommand{\ea}{\end{array}}
\newcommand{\be}{\begin{equation}}
\newcommand{\ee}{\end{equation}}
\newcommand{\bea}{\begin{eqnarray}}
\newcommand{\eea}{\end{eqnarray}}

\def\l{\lambda}

\def\d{{\rm d}}
\def\T{{\rm T}}

\newcommand{\nn}{\nonumber}
\newcommand{\eq}[1]{(\ref{#1})}

\newfont{\sgcal}{eufm9}
\newfont{\smcal}{eusm9}
\newfont{\smbf}{msbm9}
\newfont{\gcal}{eufm10}
\newfont{\mcal}{eusm10}
\newfont{\mbf}{msbm10}
\newfont{\Gcal}{eufm10 scaled\magstep1}
\newfont{\Mcal}{eusm10 scaled\magstep1}
\newfont{\Mbf}{msbm10 scaled\magstep1}
\newcommand{\der}{\partial}
\newcommand{\eps}{\varepsilon}

\def\l{\left}
\def\r{\right}

\begin{document}
\thispagestyle{empty}
\addtocounter{page}{-1}
\begin{flushright}
SNUST 01-0502\\
{\tt hep-th/0107106}\\
\end{flushright}
\vspace*{1cm}
\centerline{\Large \bf Anatomy of One-Loop Effective Action}
\vspace*{0.3cm}
\centerline{\Large \bf in}
\vspace*{0.3cm}
\centerline{\Large \bf Noncommutative Scalar Field Theories~\footnote{
Work supported in part for SJR and JTY by BK-21 Initiative in Physics 
(SNU-Project 2), KOSEF Interdisciplinary Research Grant 98-07-02-07-01-5, 
and KOSEF Leading Scientist Program, for HTS by KOSEF
Brain-Pool Program, and for YK by Korea Research Foundation.}}
\vspace*{1.2cm} 
\centerline{\bf Youngjai Kiem ${}^a$, Soo-Jong Rey ${}^{b,c}$, 
Haru-Tada Sato ${}^a$, Jung-Tay Yee ${}^b$ }
\vspace*{0.8cm}
\centerline{\it BK21 Physics Research Division \& 
Institute of Basic Science}
\vspace*{0.27cm}
\centerline{\it Sungkyunkwan University, Suwon 440-746 \rm KOREA ${}^a$} 
\vspace*{0.45cm}
\centerline{\it School of Physics \& Center for Theoretical Physics}
\vspace*{0.27cm}
\centerline{\it Seoul National University, Seoul 151-747 \rm KOREA ${}^b$}
\vspace*{0.45cm}
\centerline{\it Theory Division, CERN, CH-1211 Genev\'e \rm SWITZERLAND ${}^c$}
\vspace*{0.8cm}
\centerline{\tt ykiem, haru@newton.skku.ac.kr 
\hskip0.45cm sjrey@gravity.snu.ac.kr \hskip0.45cm jungtay@phya.snu.ac.kr}
\vspace*{1.5cm}
\centerline{\bf abstract}
\vspace*{0.5cm}
One-loop effective action of noncommutative scalar field theory with cubic
self-interaction is studied. Utilizing worldline formulation, both planar
and nonplanar parts of the effective action are computed explicitly. We find
complete agreement of the result with Seiberg-Witten limit of string 
worldsheet computation and standard Feynman diagrammatics. We prove that, 
at low-energy and large noncommutativity limit, nonplanar part of the effective
action is simplified enormously and is resummable into a quadratic action of 
scalar open Wilson line operators. 
\vspace*{1.8cm}

\begin{flushleft}
\end{flushleft}

\baselineskip=18pt
\newpage

\section{Introduction}
\setcounter{section}{1}
\setcounter{equation}{0}
\indent
Noncommutative field theories are field theories defined on noncommutative
spacetime, whose coordinates are promoted to operators:
\bea
\left[x^a, x^b \right] = i \theta^{ab},
\nonumber
\eea
and fields are multiplied in terms of the $\star$-product, 
\bea
A(x) \star B(y) := \exp_\star \left( {i \over 2} \partial_x \wedge
\partial_y \right) A(x) B(y),
\nonumber
\eea
implying nonlocal interactions.
As such, physical aspects of these theories are suspected to be significantly
different from the conventional (commutative) field theory. One of the most
significant features is the phenomenon of ultraviolet-infrared (UV-IR) mixing.
Motivated partly by the phenomenon, in \cite{ours}, we have studied the 
effective action of noncommutative scalar field theories and have found that, 
remarkably, nonplanar part of the effective action is expressible in terms 
of {\sl scalar} open Wilson line operators --- scalar counterpart of the
open Wilson lines \cite{reyunge, dasrey, grossetal, kawaietal} in 
noncommutative gauge 
theories. Specifically, for $\lambda [\Phi^3]_\star$-theory, nonplanar part 
of the one-loop effective action is given by 
\bea
\Gamma_{\rm np}[\Phi]
= {\hbar \over 2} \int {\d^d k \over (2 \pi)^d} \Phi_k[\Phi]
\, \widetilde{{\cal K}_{d \over 2}} (k \circ k) W_{-k} [\Phi],
\label{npresult}
\eea
at low-energy, large noncommutativity limit,
where 
\bea
W_k [\Phi] &:=& {\cal P}_t \int \d^d x \, 
\exp_\star \left( -g \int_0^1 \d t \vert \dot{y}(t) \vert 
\Phi(x + y(t)) \right) \star e^{i k \cdot x} \nn \\
(\Phi^n W)_k [\Phi] &:=& \left(-{\partial \over \partial g} \right)^n
W_k [\Phi], \qquad n=1, 2, 3, \cdots \qquad 
\left( g:= {\lambda \over 4 m} \right)
\label{scalarwilson}
\eea
denote the {\sl scalar} open Wilson line operators and $\widetilde{\cal K}$
represents `propagator' of the state created by the open Wilson lines. 
See Fig. 1. In \cite{ours},  much as their counterparts
in noncommutative {\sl gauge} theories \cite{reyunge, 
dasrey, grossetal, kawaietal}, we have also shown
that the {\sl scalar} open Wilson line operators
are appropriate interpolating operators for `dipoles' --- 
weakly interacting, nonlocal objects describing excitations in noncommutative
field theories. Recall that these dipoles obey so-called `noncommutative 
dipole relation': 
\bea
\Delta {x}^a = \theta^{ab} {k}_b,
\label{relation}
\eea
where ${k}_a$ and $\Delta {x}^a$ denote center-of-mass momentum and
dipole moment, respectively. As such, the dipoles are ubiquitous to any
noncommutative field theory, an aspect which would explain why the open 
Wilson line operators play prominent roles, not only in gauge theories
but also in {\sl scalar} field theories, in which neither gauge invariance 
nor gauge field is present.  

\begin{figure}[htb]
   \vspace{0cm}
   \epsfysize=6cm
   \centerline{ \epsffile{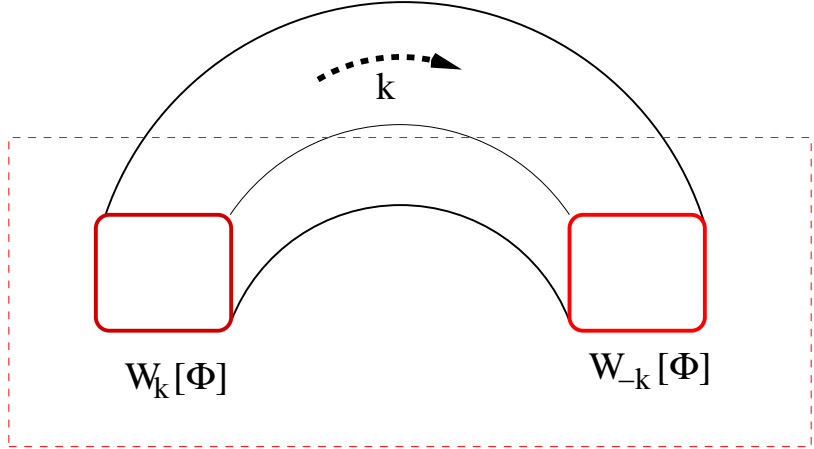} }
\caption{\sl 
Scalar open Wilson line representation of nonplanar part of the one-loop 
effective action. The open Wilson line is an interpolating field of dipole,
whose propagation is governed by the kernel ${\cal K}(k \circ k)$ in Eq.(1). }
\end{figure}

In this paper, in view of potential importance and far-reaching consequences
of the results, Eqs.(\ref{npresult}, \ref{scalarwilson}),
we present detailed computation of the one-loop effective 
action in $\lambda [\Phi^3]_\star$-theory. In \cite{ours}, the effective
action was calculated via the standard Feynman diagrammatics. To supplement
the method, in this paper, we will be computing the effective action in the 
worldline method extended to noncommutative field theories, and make detailed 
comparison, wherever possible, with methods and results in the standard 
Feynman diagrammatics as well as in the point-particle limit of the open
string worldsheet 
computation. For comparison, we compute both the planar and the
nonplanar parts of the one-loop effective action, but in low-energy, large 
noncommutativity limit. 
In this limit, as is well-known, the Weyl-Moyal correspondence enables us
to map the noncommutative
$\lambda [\Phi^3]_\star$-theory in $d$ dimensions to large-$N$, U($N$) 
matrix-valued $\lambda {\rm Tr} [\Phi^3]$-theory in $(d-2)$ dimensions. 
The large noncommutativity limit also allows us to recast the nonplanar part of 
the effective action into the form, Eq.(\ref{npresult}). 
Somewhat surprisingly, the planar part of the
one-loop effective action is not recastable in terms of closed Wilson loop
operators --- the putative `master' fields in matrix-valued field theories
at $N \rightarrow \infty$ limit.

This paper is organized as follows. 
In section 2, adopting the worldline formulation, we compute the one-loop
N-point, one-particle-irreducible Green functions of noncommutative 
$\lambda [\Phi^3]_\star$-theory. In doing so, we also observe that 
noncommutative 
vertex operators are modified into a form showing the dipole relation 
Eq.(\ref{relation}) maninfestly. 
 In section 3, we compare the result of
section 2 with open string computation of N-point S-matrix amplitude at
one-loop, and find a complete agreement in the
Seiberg-Witten scaling limit \cite{seibergwitten}.
In section 4, based on the results
in sections 2 and 3, we compute the one-loop effective action, for both 
planar and nonplanar contributions, by summing over the N-point Green 
functions. Several remarks and discussions are relegated to the last section.

Our notations are as follows.  The spacetime is taken $d$-dimensional, 
Wick rotated to Euclidean signature, with metric $G_{ab}$. 
Spacetime indices are denoted 
$a, b, c, \cdots = 1, 2, \cdots, d$. Products involving successively increasing
powers of the noncommutativity parameter $\theta^{ab}$ are denoted as:  
\bea\label{product}
p \cdot q := p_a G^{ab} q_b, \qquad
p \wedge q := p_a \theta^{ab} q_b, \qquad
p \circ q := p_a (-\theta^2)^{ab} q_b \, \cdots.
\nonumber \eea 
\section{$\lambda[\Phi^3]_\star$-Theory: Worldline Formulation}
\setcounter{section}{2}
\setcounter{equation}{0}
\indent

Begin with the worldline formulation of the noncommutative
$\lambda [\Phi^3]_\star$-theory. As stated in the Introduction, we 
are motivated to do so for detailed comparison with parallel computatons in 
the open string theory in the Seiberg-Witten limit. Moreover, the worldline 
formulation applied to noncommutative field theories, by itself, is of some 
interest
\footnote{Computations below follow closely the worldline formulation of
commutative field theories \cite{worldline}.}. 
\subsection{The effective action at one-loop}\label{sec2.1}
\indent

The classical action of the theory is given, after Wick-rotation to 
Euclidean space, by
\bea\label{act1}
S_{\rm NC} 
=\int \d^d x \, \left( {1\over2}(\der\Phi)^2 +{1\over2}m^2\Phi^2 + 
{\lambda \over3!}\Phi \star \Phi\star \Phi \,\right),
\nonumber
\eea
or, after Fourier transform, by
\bea
S_{\rm NC} =\int{\d^d k\over(2\pi)^d} \, 
{1\over2}\widetilde{\Phi}(-k)(k^2+m^2)\widetilde{\Phi}(k)
+{\lambda \over3!}\int\prod_{a=1}^3{\d^d k_a\over(2\pi)^d}
\widetilde{\Phi}(k_a) e^{-{i\over2}\sum_{i<j}k_i\wedge k_j}
(2\pi)^d \delta\left(\sum_{i=1}^3k_i \right).
\nonumber
\eea
The effective action is evaluated most conveniently by utilizing 
the background field method: split the scalar field $\widetilde{\Phi}$ as 
$\widetilde{\Phi} = \widetilde{\Phi}_0+\widetilde{\varphi}$, 
where $\widetilde{\Phi}_0$ and $\widetilde{\varphi}$ denote
classical (background) and quantum (internal) fields, respectively. 
To one-loop order, only the terms quadratic in 
$\widetilde{\varphi}$ are relevant. Explicitly, we have 
\bea\label{act3}
S_{\rm NC}&=&
\int{\d^d k_1 \over(2\pi)^d}{ \d^d k_2\over(2\pi)^d} \, 
\left[(2 \pi)^d \delta^d (k_1+k_2) {1\over2} \left(k_1^2+m^2 \right) 
\right. \nonumber\\
&+& \left. {\lambda \over2}\int{\d^d k_3\over(2\pi)^d} (2 \pi)^d 
\delta^d (k_1+k_2+k_3)
e^{-{i\over2}\sum_{i<j}^3 k_i\wedge k_j}\widetilde{\Phi}_0(k_3)
\,\right] \widetilde{\varphi}(k_1)\widetilde{\varphi}(k_2) + \cdots\ .
\eea
It shows that, compared to commutative $\lambda [\Phi^3]$-theory, 
interaction vertices are modified by Moyal's phase-factor. These phase-factors
let the scalar fields to be noncommutative while retaining associativity. We
can view the scalar fields effectively as matrix-valued fields, whose precise
nature is dictated by the so-called Weyl-Moyal correspondence map. Accordingly,
the correspondence allows us to classify Feynman diagrams in $\lambda 
[\Phi^3]_\star$-theory into the planar and the nonplanar ones \cite{filk, seiberg1, 
seiberg2}. After symmetrization 
over the momentum labelling, Eq.(\ref{act3}) is re-expressible in a form
suited for dealing with the planar and nonplanar diagrams: 
\bea\label{act4}
S_{\rm NC}&=&\int{\d^d k_1\over(2\pi)^d}{\d^d k_2\over(2\pi)^d}
(2\pi)^d \left[\, {1\over2}(k_1^2+m^2)\delta^d (k_1+k_2) \right. \nn\\
&+& \left. {\lambda \over4}\int{\d^d p\over(2\pi)^d }\delta^d (k_1 + k_2+p)
\,(e^{{i\over2}p\wedge k_1}+e^{-{i\over2}p \wedge k_1})
{\widetilde\Phi}_0(p)
\,\right]\, {\widetilde\varphi}(k_1){\widetilde\varphi}(k_2) + 
\cdots. 
\eea
Integrating out the quantum fluctuation field $\widetilde{\varphi}$, 
the one-loop effective action is given schematically as
\bea\label{Goutin}
\Gamma_{\rm 1-loop}[\Phi_0] 
= \hbar \ln\mbox{Det}^{-{1 \over 2}} \left[\, \left(k^2 +m^2 \right) 
+ {\lambda \over2}\int{\d^d p\over(2\pi)^d}
\,\left( e^{{i\over2}p\wedge k}+e^{-{i\over2}p\wedge k} \right)
{\widetilde\Phi}_0(p)\,\right]\ .
\eea
Compared to the one-loop effective action of the commutative
$\lambda [\Phi^3]$-theory: 
\bea\label{trln}
\Gamma_{\rm 1-loop}[\Phi_0] = \hbar \ln\mbox{Det}^{-{1 \over 2}}\left[\, 
\left(k^2 +m^2\right) + \lambda
\int{\d^d p\over(2\pi)^d}{\widetilde\Phi}_0(p) \,\right]\ ,
\nonumber
\eea
the interaction vertex is modified by noncommutativity in two ways. First, 
the coupling parameter is reduced effectively by a factor of 2. Its combinatoric
origin is as follows: the entire 3! diagrams can be grouped into two sets of 
3 diagrams, related one another by cyclic permutations. Due to Moyal's 
phase-factors, they constitute inequivalent diagrams. We will refer the two 
respective types of interaction vertices in Eq.(\ref{Goutin})
as ${\bf P}$ and ${\bf T}$, respectively. Second, relative sign between
${\bf P}$ and ${\bf T}$ terms is {\sl positive}. Recall that,  
for the {\sl vector} particles as in noncommutative gauge theories, the 
sign is {\sl negative}. In fact, these signs are attributed to even/odd
parity under the worldline time-reversal $\tau \rightarrow (1 - \tau)$.

For the worldline formulation, we begin with re-expressing the one-loop 
effective action Eq.\eq{Goutin} 
in path integral representation. In doing so, because of
noncommutativity in {\bf P} and {\bf T}, we will need to take care of
operator-ordering. We thus start with
\bea
-\ln\mbox{Det}\, {\cal F} (k) = \int\limits_0^\infty{\d \T\over \T} 
\hskip-6pt\int\limits_{x(\T)=x(0)}\hskip-15pt{\cal D}x(\tau)
\hskip-9pt\int\limits_{k(\T)=k(0)}\hskip-15pt{\cal D}k(\tau)
\,{\cal P}_\tau \, \exp\left( \,-\int\limits_0^\T \Big[ {\cal F} (k(\tau))- 
ik(\tau) \cdot {\dot x}(\tau) \Big] \d \tau \,\right) \ . 
\nonumber
\eea
In theories with nonderivative interactions, such as commutative 
$\lambda [\Phi^3]$-theory, the function ${\cal F}(k)$ is typically
quadratic in $k$, and hence integration over $k(\tau)$ first would be 
straightforward. In the present case, due to the $k$-dependent Moyal's 
phase-factors in {\bf P} and {\bf T}, we proceed differently and expand the 
background $\Phi_0$ first. The one-loop effective action 
then comprises of terms involving various powers of {\bf P}'s and {\bf T}'s,
in which {\bf P}$\rightarrow${\bf T} is made by an insertion of `twist' to 
adjacent internal lines. Explicitly, 
\bea\label{expG}
\Gamma_{\rm 1-loop}[\Phi_0] &=& 
{\hbar \over2}\int\limits_0^\infty {\d \T\over \T}\int\!\!\!\int
{\cal D}x {\cal D}k \,
\exp\left( \,-\int\limits_0^\T \d \tau (k^2+m^2-ik\cdot{\dot x}) \,\right)
\\
&\times&\sum_{\rm N=0}^\infty\sum_{n=0}^{\rm N} \, 
\left(-{\lambda \over2}\right)^{\rm N}\,
\Big[ \prod_{\ell =1}^n\int\limits_0^{\tau_{\ell+1}} \d \tau_{\ell}
\int{\d^d p_{\ell}\over(2\pi)^d}\,\widetilde{{\Phi}_0}(p_{\ell}) \Big]
\exp\left( \,+{i\over2}\sum_{\ell=1}^{n}
p_{\ell} \wedge k(\tau_\ell)\,\right) 
 \nn\\
&&\hskip2.5cm\times  
\Big[ \prod_{j=1}^{{\rm N}-n}\int\limits_0^{\tau'_{j+1}}\d \tau'_{j}
\int{\d^d p'_{j}\over(2\pi)^d}\, \widetilde{{\Phi}_0}(p'_{j}) \Big]
\exp\left( \,-{i\over2}\sum_{j=1}^{{\rm N}-n}
p'_{j} \wedge k(\tau'_j)\,\right) \nonumber . 
\eea
Our notation is as follows. Each summand in Eq.(\ref{expG}) comprises of 
$n$ and (N$-n$) interaction vertices with(out) twists, respectively.
See Fig. 2. 
For each group of vertices, moduli parameters are labeled as 
$\tau_\ell$ and $\tau'_j$ ($\tau_{n+1} = \tau'_{{\rm N}-n+1} = \T$), 
and external momenta are labeled as $p_{\ell}$ and $p'_{j}$, respectively.
We also assign sign factor $\nu_l=+1, -1$ to these two groups of interaction 
vertices. The two square brackets in Eq.\eq{expG} are untwisted
and twisted interaction vertices, respectively.
Therefore, for given $n$ and N, the one-loop diagram is a function
of the following set of momenta and moduli parameters:
\bea
\{\,\tau_i\,\}&=& \{\,\tau_{(l)}\quad\mbox{for}\quad i=1,2,\cdots,n\ ;\quad   
\tau'_{(j)}\quad\mbox{for}\quad i=n+1,\cdots, {\rm N} \,\} \nn \\
\{\,p_i \,\}&=& \{\,p_{(l)} \quad\mbox{for}\quad i=1,2,\cdots,n\ ;\quad   
p'_{(j)}\quad\mbox{for}\quad i=n+1,\cdots, {\rm N} \,\}\ .
\nonumber
\eea

\begin{figure}[htb]
   \vspace{0cm}
   \epsfysize=7cm
   \centerline{ \epsffile{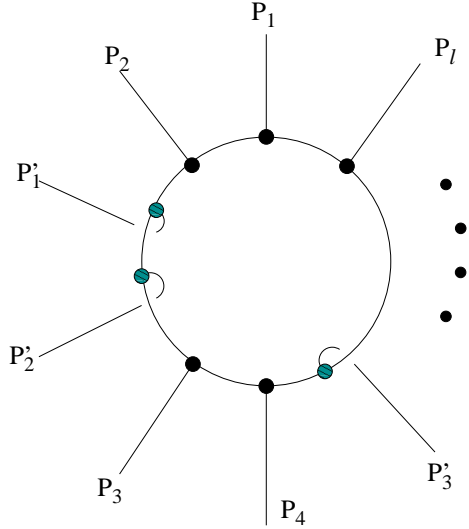} }
\caption{\sl One-Loop N-point Green function. Interaction vertices of untwisted
and twisted types are marked with solid and dashed circles, whose momenta are
labelled as $p_1, \cdots, p_\ell$ and $p'_1, \cdots, p'_{{\rm N}- n}.$}
\end{figure}

The N-point, one-particle-irreducible (1PI), Green functions are obtained
by expanding the effective action Eq.\eq{expG} in powers of $\Phi_0$. 
In the commutative setup, they are calculated by substituting the classical 
background field into a sum of `plane-wave':
\bea
\widetilde{{\Phi}_0} ( x ( \tau )) \quad \longrightarrow \quad 
\sum_{\ell = 1}^{\rm N} e^{ i p_\ell \cdot x(\tau) }.
\label{replacement}
\eea
This substitution is still valid for the present case, as
the products between 
$\widetilde{{\Phi}_0}$'s in \eq{expG} 
in the {\em momentum} represetation are local products (with explicit
Moyal's phase-factors attached). By making the plane-wave 
substitution Eq.\eq{replacement}, we now generate all the possible diagrams 
automatically (discarding terms containing the same momentum). 
This leads, for given $n$ and N, the moduli integrals in Eq.\eq{expG} to
\bea\label{rule}
&& \Big[ \prod_{\ell=1}^n\int\limits_0^{\tau_{\ell+1}} \d\tau_{\ell}
\widetilde{{\Phi}_0} \left(p_\ell \right) \Big] \cdot 
\Big[\prod_{j=1}^{{\rm N}-n} \int\limits_0^{\tau'_{j+1}} \d \tau'_{j}
\widetilde{{\Phi}_0} \left(p'_{j} \right) \Big] \nn\\
&\longrightarrow&
\prod_{\ell=1}^n   
\int\limits_0^{\tau_{\ell+1}} \d \tau_{\ell} 
\prod_{j=1}^{{\rm N}-n} 
\int\limits_0^{\tau'_{j+1}} \d \tau'_{j} 
\left[ e^{ip_1x(\tau_1)} e^{ip_2 x(\tau_2)} \cdots 
 e^{ip_{\rm N} x(\tau_{\rm N})} \nonumber \right.\\
&& \qquad \qquad \quad + (\mbox{all permutations})\, \Big]\ . 
\eea
By interchanging moduli variables $\tau$'s, all permutation terms 
in Eq.\eq{rule} can be arranged as all possible ordered integrals having
the same integrand. We find that the right-hand side of Eq.\eq{rule} 
involves the moduli-space integrals: 
\bea
\sum_{\{\nu_i\}}\int\limits_0^\T \d \tau_{\rm N} \cdots\int\limits_0^\T \d
\tau_1\,
\prod_{\ell=1}^{\rm N} e^{ip_\ell x(\tau_\ell)}\ 
=
\sum_{\{\nu_i\}} {\T \over \rm N} \int\limits_0^\T \d \tau_{\rm N-1} 
\cdots\int\limits_0^\T \d \tau_1\,
\prod_{\ell=1}^{\rm N} e^{ip_\ell x(\tau_\ell)}\ 
\nonumber
\eea
This is essentially the N-point correlator (evaluated with an 
appropriate worldline Green function). 
The combinatorics work as follows. 
The sum over $\{\nu_i\}$ takes into 
account of all possible $2^N$ terms, viz. the binomial expansion of
$({\bf P} + {\bf T})^{\rm N}$ interaction vertices.
In the commutative limit, all the $\rm 2^N$ terms reduce to the same 
contribution, and cancel eventually the $(1 / 2)^N$ factor originating
from the rescaling of the coupling parameter, $\lambda \rightarrow
\lambda / 2$. Alternatively, 
as the second second expression in the above moduli-space integral shows, 
the sum over $\{\nu_i \}$ takes into account of all possible $2^{\rm N}$-terms:
$2{\rm N}$ possibilities for the N-th reference interaction vertex, and 
$2^{\rm N-1}$ combinatoric possibilities for the rest. 
The factor of N is cancelled by the symmetry factor for the reference
vertex, 1/N. The net result is $2^{\rm N}$, yielding the same combinatoric
counting. 

Consequently, at one-loop, the $\rm N$-point Green function 
(corresponding to Eq.\eq{expG}) is given by: 
\bea\label{GN}
\Gamma_{\rm N} [p_1, \cdots, p_{\rm N}] 
&=&{\hbar \over2}\left(-{\lambda \over2}\right)^{\rm N} 
\sum_{\{\nu_i\}} \int\limits_0^\infty {\d \T\over \T}
\int\limits_0^{\rm T}\!\!\cdots\!\!\int\limits_0^\T 
\prod_{\ell=1}^{\rm N} \d \tau_\ell \, \nn\\
&\times&\int{\cal D}x \, \int{\cal D}k \, 
e^{-\int\limits_0^\T \d \tau \left[ k^2+m^2-ik \cdot 
{\dot x} \right]}
\prod_{j=1}^{\rm N} \, e^{i p_j \cdot x(\tau_j)}
   e^{ {i\over2}\nu_j p_j\wedge k(\tau_j)}\ .
\eea
A brief comment is in order. In the above derivation, for simplicity, 
we have utilized the plane-wave basis. 
As will be 
shown momentarily, the phase-factor $\exp[{i\over2}\nu_j p_j\wedge k(\tau_j)]$  
ought to be understood as part of 
a generalized vertex operator, viz. the plane-wave (scalar) vertex operator 
$\int \d \tau_\ell \, e^{i p_\ell \cdot x(\tau_\ell)}$ is not the proper one in 
noncommutative field theories. In fact, we will be showing that
the standard Feynman diagrammatics results in Appendix A 
are reproducible precisely in terms of these new vertex operators . 

\subsection{The N-point Green Function}\label{sec2.2}
\indent

We next evaluate the path integral in Eq.\eq{GN} explicitly and  
derive parametric expression for the one-loop, N-point Green function. 
In this section, we will prove that the result is given by
\bea\label{GNfinal}
\Gamma_{\rm N} [p_1, \cdots, p_{\rm N}] 
&=&{\hbar \over2}\left(-{\lambda \over2}\right)^{\rm N} 
\sum_{\{\nu_i\}}\, \int\limits_0^\infty {\d \T\over \T}\, e^{-m^2 \T}
\left({1\over4\pi \T}\right)^{d\over2}
\int\limits_0^\T \prod_{\ell=1}^{\rm N} \d \tau_\ell
\prod_{i<j}^{\rm N}  e^{{i\over2} 
\nu_{ij} \varepsilon(\tau_{ij}) p_i \wedge p_j } \nonumber\\
& &\times
\exp\left[\,{1\over2}\sum_{k, \ell =1}^{\rm N} p_k \cdot {\cal G}_{B\theta}
(\tau_k,\tau_\ell;\eps_k,\eps_\ell) \cdot p_\ell \,\,\right]\ ,
\eea
where ${\cal G}_{B\theta}^{ab}$ denotes noncommutative counterpart of 
the worldline propagator ${\cal G}_B$:
\bea\label{GBtheta}
{\cal G}^{ab}_{B\theta} \left(\tau_k,\tau_\ell;\eps_k,\eps_\ell \right)
=g^{ab} {\cal G}_B(\tau_k,\tau_\ell)-{i\over \T}
\theta^{ab}\eps_{k \ell}(\tau_k+\tau_\ell)
+{1\over4 \T}(-\theta^2)^{ab} \eps^2_{k \ell} \ .
\eea
We have introduced the following notations
\bea 
\nu_{ij} :={\nu_i + \nu_j \over2}, \quad \tau_{ij} := \tau_i - \tau_j, 
 \quad \eps_i := \frac{1- \nu_i}{2}, \quad
{\rm and} \quad \varepsilon(\tau) := {\rm sign}(\tau).
\nonumber
\eea
In addition,  $\eps_{k\ell}$ refers to $\eps_{k\ell} =
\eps_k-\eps_\ell$.  In the present case, as Eq.\eq{expG} indicates, 
the untwisted and the twisted vertices ought to be distinguished from each 
other, as the exponent of the $\star$-product flips sign, depending on 
whether the vertex is twisted or not (cf. Eq.\eq{Goutin}). One recognizes 
also that the  $\star$-product structure is dressed with twist-dependent 
`weight':
\bea
e^{ip_ix(\tau_i)}\ast^\nu e^{ip_jx(\tau_j)}
= \exp\left( \,{i\over2}\nu_{ij} \varepsilon(\tau_{ij}) 
p_i \wedge p_j \,\right)
e^{ip_ix(\tau_i)+ip_jx(\tau_j)},
\nonumber
\eea
where $\nu_{ij} = 0 , \ +1 , \  -1$ depending on types of boundaries along
which the interaction vertices are inserted.    

In the rest of this subsection, we prove Eqs.(\ref{GNfinal}, \ref{GBtheta}).
Start with the path-integral over $k(\tau)$. The relevant integral is
\bea
K := \int{\cal D}k\,\exp\left(\, -\int\limits_0^\T \d \tau \left[
k^2(\tau)-ik\cdot {\dot x}(\tau) + {i\over2}\sum_{\ell=1}^{\rm N} \nu_\ell
\delta(\tau-\tau_\ell) k(\tau)\wedge p_\ell \, \right] \right) \ ,
\nonumber
\eea
a Gaussian type integral. After shifting the momentum density as
\bea
k^a(\tau)\quad\longrightarrow\quad 
k^a (\tau)+{i\over2}{\dot x}^a (\tau)-{i\over4}
\sum_{\ell=1}^{\rm N}\delta(\tau-\tau_\ell)\nu_\ell \, \theta^{ab} 
p^b_\ell \ , 
\nonumber
\eea
the integral yields
\bea
K={\cal N}(\T) \, \exp\left(\,-{1\over4}\int\limits_0^\T \d \tau 
\left[\, {\dot x} -{1\over2} \sum_{\ell=1}^{\rm N} \theta \cdot p_\ell 
\nu_\ell \delta(\tau-\tau_\ell) \right]^2 
\,\right) \ ,
\nonumber
\eea
where ${\cal N}(\T)$ is a $\T$-dependent normalization factor, which turns
out the same as the commutative case.
Square of the $\delta$-functions in the exponent does not 
lead to divergences, as all $\tau_\ell$ take different values because of
the noncommutativity. As such, one finally finds
\bea
K={\cal N}(\T) \, \exp \left(-{1\over4}\int\limits_0^\T 
{\dot x}^2 \d \tau \right) 
\prod_{\ell=1}^{\rm N} \exp \left({1\over4}\nu_\ell
{\dot x(\tau_\ell)}\wedge p_\ell \right) .
\nonumber
\eea
Utilizing the expression $K$, the N-point correlation function is then 
reduced to 
\bea\label{GN2}
\Gamma_{\rm N}[p_1, \cdots, p_{\rm N}]
&=&{1\over2}\left(-{\lambda\over2}\right)^{\rm N} 
\sum_{\{\nu_i\}}\, \int_0^\infty 
{\d \T\over \T}e^{-m^2\T}
\int\limits_0^{\rm T} \prod_{\ell=1}^{\rm N} \d\tau_\ell \nn\\
&\times& \, {\cal N}(\T)
\hskip-5pt\int\limits_{x(0)=x(\T)}\hskip-15pt{\cal D}x\,
\exp \left( -{1\over4}\int\limits_0^\T {\dot x}^2 \d \tau \right) 
\prod_{\ell =1}^{\rm N} e^{ip_\ell x(\tau_\ell )}e^{{1\over4}{\dot x}
(\tau_\ell)\wedge 
p_\ell \nu_\ell}\ .
\eea
Next, evaluate the path integral over $x(\tau)$:
\bea
X := \int\limits_{x(0)=x(\T)}\hskip-15pt{\cal D}x\,
e^{-{1\over4}\int\limits_0^\T {\dot x}^2 d\tau}
\prod_{\ell =1}^{\rm N} \exp\left(\, i p_\ell \cdot x(\tau_\ell) 
- {\nu_\ell \over 4} p_\ell \wedge {\dot x}(\tau_\ell) \,\right)\ .
\nonumber
\eea
The integrand suggests that the vertex operator relevant for the
noncommutative 
scalar field is not the conventional plane-wave vertex operator but,
as mentioned earlier,
\bea 
V_{\rm NC} (x):=\int_0^\T \d \tau \, 
\exp\left(\, i p \cdot x (\tau) - {\nu \over 4} p \wedge {\dot x}(\tau) 
\,\right)\ . 
\nonumber
\eea
The integral $X$ is evaluated as follows. First, decompose the 
$x(\tau)$ field into normal modes:
\bea
x^\mu(\tau)=x^\mu_0\, +\, \sum_{n=1}^\infty\, x^\mu_n\, 
\sin\left({n\pi\tau\over \T}\right)\ .
\nonumber
\eea
Integral over the zero-mode $x_0$ enforces total energy-momentum 
conservation. The rest yields 
\bea
X=\int\limits_{-\infty}^\infty \prod_{n=1}^\infty \d x_n\,
\exp\left[\, -{\pi^2\over8\T}n^2x_n^2 
+ i\sum_{\ell=1}^{\rm N} p_\ell x_n\sin\left({n\pi\tau_\ell \over \T}\right)
-{1\over4}\sum_{\ell =1}^{\rm N} p_\ell \wedge x_n\nu_\ell {n\pi\over \T}
\cos\left({n\pi\tau_\ell \over \T}\right)
\,\right] . \nonumber
\eea
The $x_n$ integrations are of Gaussian type. Completing the exponent into 
squares and fixing the normalization as in the commutative case, we obtain
\bea
X = \left({1\over4\pi \T}\right)^{d\over2}
\prod_{n=1}^\infty \exp\left[\,{2\T\over n^2\pi^2}
\left(\,i\sum_{\ell=1}^{\rm N} p_\ell \sin\left({n\pi\tau_\ell \over \T}\right)
+{1\over4}\sum_{\ell=1}^{\rm N} \theta \cdot p_\ell \nu_\ell
{n\pi\over \T}\cos\left({n\pi\tau_\ell \over \T}\right)\,\right)^2
\,\right] \ . \nonumber
\eea
Applying the identities
\bea
\sin\left({n\pi\tau_i\over \T}\right)
\sin\left({n\pi\tau_j\over \T}\right)
={1\over2}\left( \,\cos{n\pi(\tau_i-\tau_j)\over \T}
-\cos{n\pi(\tau_i+\tau_j)\over \T}\,\right)\ , \quad\mbox{etc.}
\nonumber
\eea
and
\bea
&&\sum_{n=1}^\infty{\cos nx \over n^2} ={1\over4}(|x|-\pi)^2-{\pi^2\over12}
\nn\\
&&\sum_{n=1}^\infty\cos n(x-a)  =\pi\delta(x-a)-{1\over2},
\nonumber
\eea
we obtain
\bea 
X=\left(1 \over 4 \pi \T \right)^{d \over 2} \, 
\exp \Big[ \,\, &-& 
{\T \over 4} \sum_{i,j=1}^{\rm N} p_i\cdot p_j\,
\Bigl\{ \,\Bigl(\, 1-{|\tau_i-\tau_j| \over \T} \,\Bigr)^2
- \Bigl( \,1-{\tau_i+\tau_j\over \T}\,\Bigr)^2\,  
\Bigr\} 
\nonumber\\
&+& {i\over8}\sum_{i,j=1}^{\rm N} p_i\wedge p_j \nu_j \T 
  {\der\over\der\tau_j}\,
\Bigl\{\,\Bigl(\,1-{|\tau_i-\tau_j|\over \T}\,\Bigr)^2
        -\Bigl(\,1-{\tau_i+\tau_j\over \T}\,  \Bigr)^2\,
\Bigr\}
\nonumber\\
&-& 
{1\over4\T}\sum_{i,j=1}^{\rm N} p_i\circ p_j\,\eps_i\eps_j\,\,\Big]\ ,
\label{master1}
\eea
where $\delta(\tau_i\pm\tau_j)=0$ is used again. 
The differentiation with respect to $\tau_j$ in the second line
of Eq.(\ref{master1}) produces both the Filk phase-factor 
and the terms linear in $\tau$'s, which will be shown to yield precisely 
the generalized $\star$-products.   
Making use of the identities derived from the energy-momentum 
conservation: 
\bea
&&\sum_{i,j=1}^{\rm N} p_i\wedge p_j\, \nu_j\tau_i \,=\, 
 \sum_{i,j=1}^{\rm N} p_i\wedge p_j\eps_{ij}\,(\tau_i+\tau_j)\ , \nonumber\\
&&\sum_{i,j=1}^{\rm N} p_i\circ p_j\, \eps_i\eps_j \,=\, 
-{1\over2}\sum_{i,j=1}^{\rm N} \eps_{ij}^2 p_i\circ p_j  \ ,
\nonumber
\eea
we were able to arrange the $X$-integral as 
\bea
X&=&\left({1\over4\pi \T}\right)^{d\over2}
\exp\left[\, {i\over2}\sum_{i<j}^{\rm N}
p_i\wedge p_j \nu_{ij}\eps(\tau_{ij})\,\right]\nn\\
&\times& \exp\left[\,\,
{1\over2}\sum_{i,j=1}^{\rm N}p_i\cdot p_j\, {\cal G}_B(\tau_i,\tau_j)
-{i\over2 \T}\sum_{i,j=1}^{\rm N}p_i\wedge p_j\eps_{ij}\,(\tau_i+\tau_j)
+{1\over8 \T}\sum_{i,j=1}^{\rm N} \eps^2_{ij}\,p_i\circ p_j\,\,\right].
\nonumber
\eea
Putting the above result into Eq.(\ref{GN2}), we finally obtain
the aforementioned expression, Eq.\eq{GNfinal}, for the N-point Green
functions at one-loop order.

The N-point Green functions, Eq.(\ref{GNfinal}), can also be obtained from
rearranging the standard Feynman diagrammatics. This is elaborated in 
Appendix A.  In comparing the two results, one should exercise a caution 
that, upon reversing the orientation of the underlying string worldsheet, one
also flips the overall sign of the phase-factor in Eq.\eq{GNfinal}. In
fact, the the overall sign choice is fixed only after the orientation 
convention is chosen.

\section{Comparison with String Worldsheet Computation}\label{sec4}
\setcounter{section}{3}
\setcounter{equation}{0}
\indent

Having found the N-point Green functions in the worldline formalism, 
we now compare Eqs.(\ref{GNfinal}, \ref{GBtheta}) with those obtained from
the string theory computation \cite{string}.   At one-loop, the relevant string worldsheet 
diagram is an annulus with two boundaries. We will parametrize the worldsheet
by complex coordinate, $z = x + i y$, where $y$ is periodically
identified as $y \simeq y + t$ and the two boundaries are located at 
$x = 0$ ($\eps = 1$ and $\nu = -1$, the inner boundary) and 
$x = 1/2$ ($\eps = 0$ and $\nu =1$, the outer boundary), respectively.  
Here $t$ is the annulus modulus.  External open strings can be 
inserted along either of the two 
boundaries, a direct counterpart of the twisted and untwisted interaction
insertions in one-loop Feynman diagrammatics.  

\begin{figure}[htb]
   \vspace{0cm}
   \hspace{-5cm}
   \epsfysize=6cm
   \centerline{ \epsffile{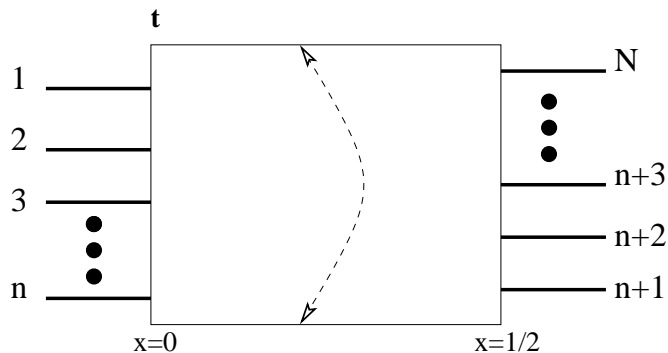} }
\caption{\sl Annulus as one-loop string worldsheet diagram. Tachyon vertex
operators are inserted on either boundary of the annulus. In the Seiberg-Witten limit, the annulus size modulus $t$ is scaled to infinity.}
\end{figure}

As we want to extract 
information concerning noncommutative {\sl scalar} field theories,
the relevant external string states are those of tachyons, whose
vertex operator is given by
\begin{eqnarray}
 V_T (p, y) = g_s \sqrt{\alpha^{\prime}} e^{ i p \cdot X(y) }  ~ ,
\nonumber
\end{eqnarray}
and we turn on the constant background two-form gauge fields, which 
turns itself into the noncommutativity parameter $\theta$ in the Seiberg-Witten
limit. The relevant N-point tachyon S-matrix amplitude, which is 
depicted in Figure 3, is  schematically expressible as (up to normalization) 
\begin{eqnarray}
{\cal A} & = & \int_0^{\infty} \frac{\d t}{t} Z(t) \int_0^t \d y_1 \cdots
\int_0^t \d y_{\rm N} \Big< V_{T 1} ( p_1 , y_1 ) \cdots 
 V_{T {\rm N}} ( p_{\rm N } , y_{\rm N} ) \Big>_t   \nonumber \\
& = &  ( g_s^2  \alpha^{\prime} )^{ \frac{\rm N}{2}} 
   \int_0^{\infty} \frac{\d t}{t} Z(t) \int_0^t \d y_1 \cdots
\int_0^t \d y_{\rm N}  \exp \left( - \alpha^{\prime} 
 \sum_{i < j}^{\rm N}   p_i { \cal G }^{ij}  p_j \right) ~ ,
\nonumber
\end{eqnarray}
in terms of the worldsheet partition function $Z(t)$ and the
worldsheet Green function ${\cal G}^{ij}$.  In the case of the annulus
partition function, nonzero two-form $B_{mn}$ background does not 
change the worldsheet Green function, except that the metric is replaced,
in the Seiberg-Witten limit, by the open string metric $G_{ab}$:
\bea
Z(t) = \int \frac{\d^{d} k}{(2 \pi )^{d}} \sum_{\{\rm I\}}  
 e^{- 2 \pi \alpha^{\prime} t ( k \cdot G \cdot k + M_{\rm I}^2 ) } 
 = \left( {1 \over 2 \pi \alpha^{\prime} t} \right)^{d \over 2} 
\, f_1 (q )^{-24} ~ ,
\nonumber
\eea
where the summation $\{\rm I\}$ is over the entire string
states in the intermediate channel, and  
\bea
f_1 (q) = q^{1/24} \prod_{m=1}^{\infty} ( 1 - q^m ) \qquad
{\rm where} \qquad q = e^{ - 2 \pi t} ~ ,
\nonumber
\eea
as are relevant for the bosonic $D_{(d-1)}$-branes.

The boundary worldsheet propagator has the following form \cite{klp}:
for two points on the same boundary, relevant for the planar diagram
contributions, 
\begin{equation}
\label{pprop1}
{\cal G}^{ab}_{\rm p} (z_i , z_j ) =  \alpha^{\prime}
   G^{ab} {\cal G} (z_i , z_j ) + \frac{i}{2}
\theta^{ab} \varepsilon (z_i -z_j ),
\end{equation}
while for two points on different boundaries, relevant for the nonplanar 
diagram contributions,
\begin{equation}
{\cal G}^{ab}_{\rm np} (z_i , z_j ) = \alpha^{\prime}
   G^{ab} {\cal G}(z_i , z_j ) + \displaystyle
\frac{(\theta \cdot G \cdot \theta )^{ab}}{2\pi \alpha^{\prime} t}  
(x_i -x_j )^2  - \frac{2 i}{t} \theta^{ab} (x_i -x_j )(y_i + y_j ).
\label{npprop1}
\end{equation}
Here, the function ${\cal G}(z_i , z_j )$ is defined by
\bea
{\cal G}(z_i ,z_j ) =  -\log \left|
\frac{\vartheta_1 (z_i -z_j |it)}{\vartheta'_1(0|it)} \right|^2 +
\frac{2\pi}{t} (y_i -y_j )^2 ~,
\nonumber
\eea
in terms of the theta function $\vartheta_1$.  

To extract the noncommutative scalar field theory amplitudes from 
open string tachyon S-matrix amplitudes, we will take the Seiberg-Witten 
decoupling limit:  
$\alpha^{\prime} \rightarrow 0$ under which the massive string modes decouple, 
while open string metric $G^{ab}$ and noncommutativity $\theta^{ab}$ 
are held 
fixed.  In fact, technically speaking, what we really do here is to 
isolate the loop contribution from the tachyon intermediate state.  
This contribution is exponentially 
divergent and dominates contributions from higher mass 
intermediate states.  We then
analytically continue the mass parameter $m^2$ to a proper positive value
to match our cubic field theory mass paramenter.  
In this process, we also keep the field theory moduli parameters 
$\T$ and $\tau$ fixed by putting
\bea
 2 \pi \alpha^{\prime} t = \T  \qquad {\rm and} \qquad
 2 \pi \alpha^{\prime} y = \tau ~ ,
\nonumber
\eea
viz. the annulus becomes infinitely thin, making essentially a circle,
relevant for one-loop Feynman diagram.  Through this procedure, one finds
that the partition function $Z(t)$ turns into
\bea
 Z(t) \rightarrow  \left( \frac{1}{ \T } \right)^{\frac{d}{2}}
                   e^{ - m^2 \T } ~ ,
\nonumber
\eea
matching the corresponding factor in the field theory result, 
Eq.(\ref{GNfinal}).  

The two-point function ${\cal G}$ in the decoupling limit
is reduced to (see for instance \cite{rolandsato}):  
\bea
 - \alpha^{\prime} {\cal G} (z_i , z_j) \longrightarrow
   {\cal G}_B = | \tau_i - \tau_j | - \frac{ ( \tau_i - \tau_j )^2}{\T} ~ ,
\nonumber
\eea
viz. only zero-mode part of $\vartheta_1$ remains. Also, noting
that $( x_i - x_j ) = -\eps_{ij} /2$ vanishes when the $i$ and $j$-th 
insertions are on the same boundary, the last two terms in the nonplanar
propagator Eq.\eq{npprop1} are reduced to the last two terms in 
Eq.\eq{GBtheta}. Likewise, the second term in the planar propagator 
Eq.\eq{pprop1} gives rise to the Filk phase-factors, as, when  $i$ and 
$j$-th insertions are along the same boundary, 
$\eps ( z_i -  z_j ) =  \eps ( \tau_{ij} ) $ at $x = 0$ ( $\nu = -1$ ) and 
$\eps ( z_i -  z_j ) = - \eps ( \tau_{ij} ) $ at $x = 1/2$ ($\nu = 1$).
Putting these observations
together, we conclude that Eq.(\ref{GNfinal}) and Eq.(\ref{GBtheta}) 
follows precisely from the string theory 
computation in the Seiberg-Witten limit.  

The expression Eq.(\ref{GNfinal}) is the general one
for a given value of N, the total number of external
scalar insertions; the sum over ${\{\nu_i\}}$ is over
$2^{\rm N}$ possible terms, spanning the cases of inner or outer
boundary insertion for each interaction vertex.  Decomposing
$\rm N = N_1 + N_2$ where $\rm N_1$ is the number of inner boundary
insertions and $\rm N_2$ is the number of outer boundary insertions,
the terms of Eq.(\ref{GNfinal}) can be classified into planar and nonplanar
contributions, depending on the value of $\rm N_1$: two terms, 
$\rm N_1 = 0$ or N, are planar diagrams, while $\rm 0 < N_1 < N$ are 
nonplanar diagrams (consisting of $\rm N! / ( N_1 ! N_2 !)$
symmetrization of the external momenta). The nonplanar diagrams correspond 
to the double trace terms 
\bea
{\rm Tr}  \underbrace{\Phi (p_1)  \cdots 
 \Phi (p_{\rm N_1} ) }_{\rm N_1}   {\rm Tr} 
\underbrace{\Phi  (p_{\rm N_1 +1} ) \cdots \Phi ( p_{\rm N} ) }_{\rm N_2} ~. 
\nonumber
\eea
For fixed $\rm N_1$, let us denote the net momentum
flow between the inner and the outer boundaries as
\bea
 P = \sum_{i=1}^{\rm N} \eps_i p_i = \sum_{r=1}^{\rm N_1} p_r ~ .
\label{transfermomentum}
\eea
From here on, the indices $r, s, \cdots$ runs from $1$ to
$\rm N_1$ (inner insertions) while the indices $m, n, \cdots$ 
runs from $0$ to $\rm N_2$ (outer insertions).  Using the
the overall momentum 
conservation, we find that the contribution to the amplitudes
from the third term of Eq.(\ref{GBtheta}) can be written as
\bea 
\exp \left(  - \frac{P \circ P }{4\T} \right) ~ . 
\nonumber
\eea
Another useful identity that can also be proved using the
momentum conservation is
\bea
\frac{1}{2} \sum_{i,j=1}^{\rm N} p_i \wedge p_j \eps_{ij}
( \tau_i + \tau_j ) = \frac{1}{2} \sum_{i<j}^{\rm N} 
 p_i \wedge p_j (\nu_i + \nu_j ) \tau_{ij} ~ .
\nonumber
\eea
The quantity $( \nu_i + \nu_j ) /2$ equals to $+1$ when
$i$ and $j$ are both outer insertions, $-1$ when
they are both inner insertions, and $0$ otherwise.  
Thus, for fixed $\rm N_1$,  each term in Eq.(\ref{GNfinal}) can be 
expressed as
\begin{eqnarray}
\Gamma_{{\rm N}, \{ \nu_j \} } &=&
{\hbar\over2} \left(-{\lambda \over2}\right)^{\rm N} 
\int_0^\infty {\d \T\over \T}\left({1\over4\pi \T}\right)^{d\over2} 
\T^{\rm N_1+ N_2} \exp \left[ - m^2 \T - \frac{P \circ P}{4\T} \right] \nn \\ 
&&\times
\left( \prod_{r=1}^{\rm N_1}\int\limits_0^1 d\tau_r\right)
\exp \left( - {i\over2}\sum_{r<s }p_r\wedge p_s \eps(\tau_{rs}) 
  +  i p_r \wedge p_s \tau_{rs} \right) \nn\\
&&\times 
\left(\prod_{a=1}^{\rm N_2}\int\limits_0^1 d\tau_a\right) 
\exp \left(  + {i\over2}\sum_{a<b}p_a\wedge p_b \eps(\tau_{ab}) 
   - i p_a \wedge p_b \tau_{ab} \right)\nn\\
&&\times \exp \left ( \T \sum_{i<j} p_{ i a} G^{ab} p_{j b}  
 \left( | \tau_i - \tau_j | - ( \tau_i - \tau_j )^2 \right) 
 \right) ~ , 
\label{wow}
\end{eqnarray}
where we have rescaled $\tau$'s by T so that they take values in the
interval $[0, 1]$. The amplitude expression Eq.(\ref{wow}) is
essentially identical to four point amplitudes ($\rm N_1 + N_2 = 4$) 
obtained in \cite{liumichelson} in the case of the ${\cal N} =4$ 
noncommutative $U(1)$ gauge theory up to a number of details.
First, the polarization dependence of the gauge fields
are deleted in the scalar field theory case.  Secondly, 
$(-1)^{{\rm N}_1}$ factor is absent reflecting the difference in
the parity under $\tau \rightarrow - \tau $ between the tachyon 
vertex operator (with even worldsheet 
oscillation number) and the gauge vertex operator (with odd
worldsheet oscillation number).  Third, while we had to rely 
on the analytic continuation to make the $m$ value an appropriate number
for the scalar theory, one can rely on Higgs mechanism, i.e., 
the separation $r$ between two parallel D-branes, to produce the mass 
$ m = r / (2 \pi \alpha^{\prime} )$ for the gauge theories.  
One further notes that the summation over $\rm N! / (N_1 ! N_2 ! )$
terms fully symmetrizes the external momenta for each nonplanar
amplitude, in line with the symmetric trace prescription
in nonabelian Born-Infeld theory.    

To get further insight into the amplitude Eq.(\ref{wow}), we now 
expand the last line in Eq.(\ref{wow}) at low-energy limit:
\bea
p_{i a} G^{ab} p_{j b} \ll m^2 \qquad{\rm for}\,\,\, {\rm every} \,\,\,
i , \, j.
\nonumber
\eea
In this limit, being subdominant compared to the first line, the last line in 
Eq.\eq{wow} simply drops out. The leading term in 
this expansion exhibits factorization property \cite{klp2} manifestly: 
\begin{eqnarray}
\Gamma_{\rm N, \{ \nu_j \} } &=&
{\hbar \over2} \left(-{\lambda \over2}\right)^{\rm 
N} \int_0^\infty {\d \T\over \T}\left({1\over4\pi \T}\right)^{d \over 2} 
\T^{\rm N_1+ N_2}
  \exp \left[ - m^2 \T - \frac{P \circ P}{4 \T} \right] \nn \\ 
&&\times
\left( \prod_{r=1}^{\rm N_1}\int\limits_0^1 d\tau_r\right)
\exp \left( - {i\over2}\sum_{r<s }p_r\wedge p_s \eps(\tau_{rs}) 
  +  i p_r \wedge p_s \tau_{rs} \right) \nn\\
&&\times 
\left(\prod_{a=1}^{\rm N_2}\int\limits_0^1 d\tau_a\right) 
\exp \left( + {i\over2}\sum_{a<b}p_a\wedge p_b \eps(\tau_{ab}) 
   - i p_a \wedge p_b \tau_{ab} \right) ~ . \label{wow1}
\end{eqnarray}
The effective action is then obtained by computing the moduli parameter
integrals explicitly and then summing over $\Gamma_{{\rm N}, \{\nu_j\}}$ 
along with the combinatoric weight $1/{\rm N}!$, as explained above
Eq.(\ref{GN}). We elaborate the details in the next section. 
\section{Effective Action, $\star_n$-products and open Wilson lines}
\setcounter{section}{4}
\setcounter{equation}{0}
\indent

Begin with evaluation of the moduli parameter integrals of the factorized
low-energy expression, Eq.(\ref{wow1}).
As elaborated in the previous section, by rescaling the vertex position 
moduli $\tau$'s by $\T \cdot \tau$, the moduli integrals in T, $\tau_r, 
\tau_a$ are also factorized. As such, we evaluate first the T-modulus integral.
Recall that the $\T$-modulus corresponds, in the open 
string worldsheet computation, 
to the modulus of annulus diagram. One readily obtains
\bea
{\cal K}_{\rm N} \left( P, \Lambda; d \right)
&:=& 
\int\limits_0^\infty {\d \T \over \T} 
\, \left({1 \over 4 \pi {\rm T}} \right)^{d \over 2}
\T^{\rm N} \exp\left[-m^2 \T - { P \circ P + \Lambda^{-2} \over 4 \T} 
     \right]   \nonumber \\
   &=& \left( {1 \over 2 \pi} \right)^{d \over 2} 2^{- \rm N - 1} 
\l( { P \circ P + \Lambda^{-2} \over m^2}  \r)^{{\rm N\over 2} -{d \over 4}}
     {\bf K}_{{\rm N} - {d \over 2} } \Big( m \vert P \circ P + 
\Lambda^{-2}  \vert^{1 \over 2} \Big) ~ ,
\label{bess}
\eea
where we have introduced the UV cutoff $\Lambda$ explicitly, and dependence
on $P, \Lambda$, and spacetime dimension $d$ are emphasized. The function 
$ {\bf K}_{{d \over 2}-{\rm N}}(z)$ refers to the modified Bessel
function.   For the planar diagrams, inferred from 
Eq.(\ref{transfermomentum}), $P = 0$ and hence the UV cutoff $\Lambda$ is 
indispensible.  

Next, evaluate the moduli integrals in the second and the third lines 
in Eq.(\ref{wow1}). These integrals turn out to be identical to the definition of 
generalized $\star_{\rm N}$-products, as defined, for instance, in 
\cite{liustar}.
Note that we have decomposed N-point interaction vertices into 
$\rm N_1$ untwisted ones and $\rm N_2$ twisted ones, where 
$\rm N = N_1 + N_2$.  In the string worldsheet computation, the former
type of insertions corresponds to the `outer' boundary 
insertions, and the latter to the `inner' boundary insertions.  
One readily finds that each group of the insertions yields a cluster of 
the generalized $\star$-products. The T-integral, ${\cal K}_{\rm N}
(P, \Lambda; d)$, 
then supplies a sort of `propagator', connecting the two clusters
of generalized $\star$-products. See figure 1.

As emphasized already, the generalized $\star$-product arises when there
exists a net momentum flow between the two clusters of external lines,
viz. between untwisted and twisted interacton vertices. As denoted in 
Eq.(\ref{transfermomentum}), the net momentum flow $P$ is given by
\bea
 P + p_1 + \cdots + p_{\rm N_2} = 0 ~ .
\nonumber
\eea
Making use of the identity
\bea
 \sum_{a<b=1}^{\rm N_2} p_a \wedge p_b ( \tau_a - \tau_b ) 
 = \sum_{a=1}^{\rm N_2} P \wedge p_a \tau_a ~ ,
\nonumber
\eea
we can re-express the third line of Eq.(\ref{wow1}) as
\begin{eqnarray}
& & \left(\prod_{a=1}^{\rm N_2}\int\limits_0^1 d\tau_a\right) 
\exp \sum_{a<b=1}^{\rm N_2}\left(  {i\over2}\eps(\tau_{ab}) \,
p_a\wedge p_b - i  \tau_{ab} p_a \wedge p_b \right)  \nonumber \\
& =&  \left(\prod_{a=1}^{\rm N_2}\int\limits_0^1 d\tau_a\right) 
\exp \sum_{a<b=1}^{\rm N_2}
\left(  {i\over2} \eps(\tau_{ab})~ p_a \wedge p_b \right) 
\exp \left( - i \sum_a P \wedge  p_a \tau_a \right) ~ .
\label{usefst} 
\end{eqnarray}
As expressed, the moduli integrals over $\tau$'s are unordered and 
ranges over the entire circle $[0, 1]$. One can decompose these integrals
into $\rm N_2 !$ ordered integrals, each of which is defined with a definite
ordering among the $\rm N_2$ $\tau_a$-moduli parameters. For each ordering,
the first exponential in Eq.(\ref{usefst}) gives rise to Filk's phase-factor, which, in the absence of the second exponential, simply yields 
the standard $\star$-product.  In the case of the planar contribution, $P = 0$
and the relevant product is the {\em symmetrized} form of the standard 
$\star$-product:
\bea
 \left[ A_1 A_2 \cdots A_{\rm N} \right]_{\star_{\rm sym}}
 := \frac{1}{\rm N!} \sum_{\{\rm perm\}} A_{i1} \star \cdots  \star 
  A_{i{\rm N}} ~ , 
\nonumber
\eea
where the summation is over ${\rm N}!$-permutations. In the case of 
the nonplanar contributions, however, because of nonvanishing momentum
flow $P$, the relevant product turns out to be the generalized 
$\star_{\rm N}$-product \footnote{Relevance of generalized $\star$-products 
and relation to gauge invariance and Seiberg-Witten map have been studied 
recently \cite{gaugegeneralizedstar}, but all in the context of noncommutative 
{\sl gauge} theories.}.

The explicit evaluation of Eq.(\ref{usefst}), including the 
nonabelian Chan-Paton factor, was made in \cite{klp2}.
The results are:
\begin{eqnarray}
{\rm Tr } \left[ f_1 (p_1 ) , f_2 (p_2) , \cdots
 f_{\rm N_2} ( p_{\rm N_2} ) \right]_{\star_{\rm N_2}} &=&  
\sum_{\rm (N_2 -1)! } f^{a_1}_1 (p_1 )
 \cdots f_{\rm N_2}^{a_{\rm N_2}} (p_{\rm N_2} ) \,
 {\rm Tr} \left( {\tt T}^{a_1} \cdots {\tt T}^{a_{\rm N_2}} \right) 
\nonumber \\
&\times& \left( \frac{ \exp \left[  \frac{i}{2}
 \sum\limits_{i<j}^{\rm N_2} p_i \wedge p_j \right]}
 { \prod_{i=2}^{\rm N_2} \left( -i k \wedge P_i \right) }
 + ({\rm cyclic} ~ {\rm permutations } ) \right),
\nonumber
\end{eqnarray}
where the summation runs over the $\rm (N_2 -1 )!$ noncyclic
permutations (with independent Chan-Paton factor),
$f_i = \sum_{a_i} f_i^{a_i} {\tt T}^{a_i}$, ${\tt T}^{a_i}$ are generators of
the $U(n)$ Chan-Paton group, and 
$P_i := \sum_{j=i}^{\rm N} p_{j}$. For $U(1)$ gauge group, they reduce to:
\begin{eqnarray}
\left[ f_1 (p_1 ) , f_2 (p_2) , \cdots
 f_{\rm N_2} ( p_{\rm N_2} ) \right]_{\star_{\rm N_2}} &=& 
 \sum_{\rm (N_2 -1)! } f_1 (p_1 )
 \cdots f_{\rm N_2}  (p_{\rm N_2} ) \nonumber \\
 &\times& \left( \frac{ \exp \left[  \frac{i}{2}
 \sum\limits_{i< j}^{\rm N_2} p_i \wedge p_j \right]}
 { \prod_{i=2}^{\rm N_2} \left( -i k \wedge P_i \right) }
 + ({\rm cyclic} ~ {\rm permutations } ) \right) ~ . 
\nonumber
\label{u1result}
\end{eqnarray}
One can explicitly work out and find that 
they are given by
\begin{equation}
\left[A(x_1) B(x_2) \right]_{\star_2} 
:= { \sin \left({1 \over 2} \partial_1 \wedge \partial_2 \right) 
\over {1 \over 2} \partial_1 \wedge \partial_2} A(x_1) B(x_2) 
\nonumber
\end{equation}
\begin{eqnarray}
\left[ A(x_1) B(x_2) C(x_3) \right]_{\star_3}
:=
   \left[ {\sin \left( {1 \over 2} \partial_2 \wedge \partial_3 \right)
\over {1 \over 2}(\partial_1 + \partial_2) \wedge \partial_3 }
{\sin \left({1 \over 2} \partial_1 \wedge (\partial_2 +\partial_3) \right)
\over {1 \over 2} \partial_1 \wedge (\partial_2 +\partial_3)}
+(1 \leftrightarrow 2) \right] A(x_1) B(x_2) C(x_3)
\nonumber
\end{eqnarray}  
and so forth.  Evidently, as $k$ subset of momenta go to zero, $\star_{\rm N}$
is reduced to $\star_{{\rm N} - k}$.

Combining Eq.(\ref{bess}) and Eq.(\ref{u1result}), Eq.(\ref{wow1}) can
be rewritten as
\begin{equation}
\Gamma_{\rm N, \{ \nu_j \} } = \frac{\hbar}{2} 
\left( - \frac{\lambda}{4} \right)^{\rm N_1 + N_2 } 
\left[\Phi \cdots\Phi \right]_{\star_{\rm N_1}} \left( - P \right) 
\widetilde{{\cal K}}_{{\rm N}- d/2} \left( P \circ P
  + 1/ \Lambda^2  \right) 
 \left[ \Phi \cdots\Phi \right]_{\star_{\rm N_2}} (+ P )  ~ ,
\label{starexpr}
\end{equation}
where the kernel ${\cal K}_n$ is given, from Eq.(\ref{bess}), by
\bea
  \tilde{{\cal K}}_n ( z^2 ) = 
 2 \left( \frac{1}{2 \pi } \right)^{d \over 2} 
 \left( \frac{ |z| }{m } \right)^n  {\bf K}_{n} 
 \left( m | z | \right) .
\nonumber
\eea
In Eq.(\ref{starexpr}), we have retained the UV cutoff $\Lambda$, as the
result is equally valid for planar contributions in so far as $P$ is set to 
zero and the generalized $\star$-products are replaced by the standard 
$\star$-products.

As is clear from the defintion in Eq.(\ref{u1result}), the generalized star
products are symmetric with respect to the external momenta.
Hence, each summand in the $2^{\rm N_1 + N_2}$-summations over 
$\left\{ \nu_j \right\}$ yield the same contribution as long as $\rm N_1$ 
(thus $\rm N_2$) is the same; this gives rise to the combinatoric factor 
$\rm N!/(N_1 ! N_2 !)$.  Recalling the definition of the effective action 
in our convention with the $\rm 1/N!$ factor, we finally obtain the 
one-loop effective action as 
\begin{eqnarray}
\Gamma[\Phi] &=& 
\frac{\hbar}{2} \sum_{\rm N=1}^\infty 
\left( - \frac{\lambda}{4} \right)^{\rm N}
\frac{1}{\rm N!} \int \d^d x 
\sum_{\rm N_1 = 0}^{\rm N} \frac{\rm N!}{\rm N_1! ( N- N_1 )! } \nonumber \\ 
&& \times
\left[ \Phi \cdots \Phi \right]_{\star_{\rm N_1}} ( x ) ~ 
\widetilde{{\cal K}}_{{\rm N} - d/2} \Big(- \partial_x  \circ \partial_x
  + \Lambda^{-2}  \Big) ~ 
 \left[ \Phi \cdots \Phi \right]_{\star_{\rm N - N_1}} ( x )  ~,
\nonumber
\end{eqnarray}
encompassing both planar ($\rm N_1 = 0$ or N) and nonplanar contributions.
As we will see below, the planar contribution is the counterpart of the
closed
Wilson loop one-point function, viz. counterpart of the closed string tadpole, 
while the nonplanar contribution is the counterpart of the open Wilson line
two-point functions. See figure 4.

\begin{figure}[htb]
   \vspace{0cm}
   \epsfxsize=14cm   
   \epsfysize=5cm
   \centerline{ \epsffile{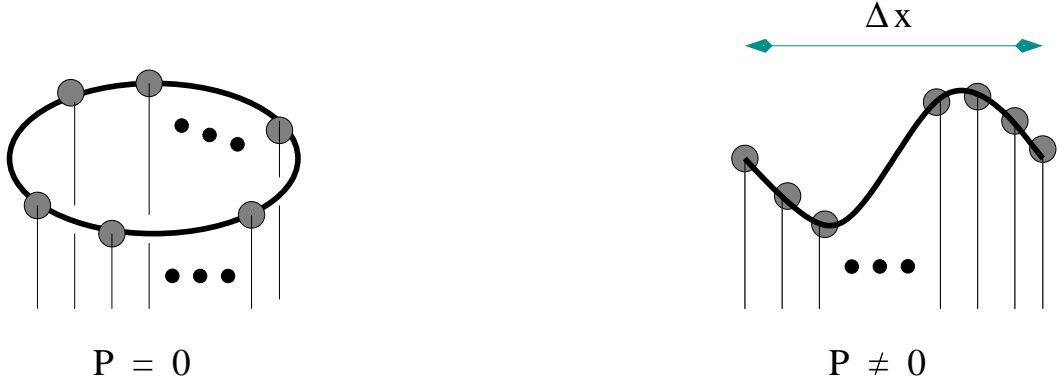} }
\caption{\sl Spacetime snapshot of planar and nonplanar contributions. 
Partial momentum sum of (un)twisted interaction vertices is denoted as $P$.
For $P=0$, viz. planar contribution, the virtual quanta sweeps a closed path 
in spacetime. When $P \ne 0$, viz. nonplanar contribution, the virtual 
quanta jumps $\Delta x$ in spacetime. }
\end{figure}

\subsection{Planar Contribution}
We first consider the planar contribution $\rm 
N_1 = 0$ or N in the limit where 
the momentum cutoff is much larger than the mass scale $m$, $\Lambda \gg m$.  
From the Taylor expansion of the modified Bessel function,
we obtain the following expressions for the kernel:
\bea
  {\cal K}_n ( z^2 ) = 
 2 \left( \frac{1}{2 \pi } \right)^{d/2} 
 \left( \frac{ |z| }{m } \right)^n  \frac{\Gamma ( | n | )}{2}
  \left( \frac{2}{m |z| } \right)^{|n|}   
\nonumber
\eea
for $n \ne 0$ and
\bea
  {\cal K}_0 ( z^2 ) = 
 2 \left( \frac{1}{2 \pi } \right)^{d \over 2} 
   \left( - \log \frac{ m |z| }{2} 
  \right) 
\nonumber
\eea
for $n= 0$ and $n= {\rm N} - d/2$.  Here, $z =  1/ \Lambda$, and, as explained
above, the generalized star products should be understood as the standard
star products: 
\begin{eqnarray}
\left[ \Phi, \cdots, \Phi \right]_{\star_{\rm N_1}} ( x ) 
 \left[ \Phi, \cdots, \Phi \right]_{\star_{\rm N - N_1}} ( x ) =
  \left[ \Phi \star \cdots \star \Phi \right] (x) ~ ,
\nonumber
\end{eqnarray}
viz. N-tuple of the standard $\star$-products, etc. Details of the UV divergence
depends on the spacetime dimension $d$. For instance, in $d=6$ wherein the
theory is renormalizable, the two-point Green function ($\rm N=2$) is 
quadratically divergent, and the three-point Green function ($\rm N=3$) 
is logarithmically divergent. 
The higher-point functions ($n>0$) are finite, as dependence on the UV cutoff 
$\Lambda$ cancels out. One furthermore observes that, 
after renormalization of the divergent contributions, the combinatorics 
factors come out as
\bea
 \frac{\rm (N-4)!}{\rm N!} = \frac{1}{\rm N (N-1)(N-2)(N-3) }
 = -\frac{1}{6} \left( {1 \over \rm N} - \frac{3}{\rm N-1} + 
\frac{3}{\rm N-2} - \frac{1}{\rm N-3} \right) ~ .
\nonumber
\eea
As such, the planar contribution to the effective action is given by:
\begin{equation}
\Gamma_{\rm p} \simeq  \hbar 
\left( m^2 + \frac{\lambda}{2}  \Phi \right)^3_\star \star \log_\star 
\left ( m^2 + \frac{\lambda}{2} \Phi \right).
\label{finalpl}
\end{equation}
The result Eq.(\ref{finalpl}) is precisely noncommutative version of 
the `Coleman-Weinberg'-type potential, where the parameters and the fields are
to be understood as renormalized ones. 

The planar contribution ought to correspond, in string theory, to the diagrams
with a tadpole insertion \cite{tadpole}. 
This is evidently so, except one puzzling point: in the large noncommutativity
limit, the Weyl-Moyal correspondence permits to map the noncommutative
$\lambda [\Phi^3]_\star$-theory into large-$N$ limit of U($N$) matrix 
$\lambda {\rm Tr}[\Phi]^3$-theory.
One would have expected that the dominant dynamics is describable 
in terms of the standard Wilson loop operators
\bea
W_0[\Phi] := \int \d^4 x \, \exp_\star \left(- \lambda \Phi(x) \right),
\nonumber
\eea
where the integration is over the noncommutative directions.
Typically, these Wilson loops are the large-$N$ limit `master' fields in 
matrix-valued field theories. Apparently, the result, Eq.(\ref{finalpl}), 
does not involve the above Wilson loops, even after taking the large 
noncommutativity limit. Whether this discrepancy invalidates the concept 
of master field in this context or not is unclear yet. 
\subsection{Nonplanar Contribution}
The behavior of the nonplanar contribution is markedly different
from those of the planar part, especially as we take the
large noncommutativity limit.  From here on, we will drop the
cutoff by sending it to infinity and Wick-rotate back to the
Minkowski space.  The nonplanar part of the effective action
then becomes a double sum involving the generalized $\star$-products:
\bea
\Gamma_{\rm np} = {\hbar \over 2}
\sum_{\rm N=2}^{\infty} \l( - { \lambda \over 4}\r)^{\rm N} 
  {1 \over {\rm N}!}
\int \d^d x  \sum_{n=1}^{\rm N-1} {{\rm N} \choose n} 
\Big[\Phi \cdots \Phi \Big]_{\star_{\rm n}} (x) \,
{\cal K}_{{\rm N} - {d \over 2}}\left( - \partial_x \circ \partial_x \right) 
\, \Big[\Phi \cdots \Phi \Big]_{\star_{{\rm N} - n}}(x). 
\nonumber
\eea

To proceed further, we will be taking the low-energy, large noncommutativity 
limit:
\bea
q_\ell \sim \epsilon, \qquad
{\rm Pf} \theta \sim {1 \over \epsilon^2} \qquad {\rm as}
\qquad \epsilon \rightarrow 0
\label{limit1}
\eea
so that
\bea
q_\ell \cdot q_m \sim {\cal O} (\epsilon^{+2}) \rightarrow 0, 
\qquad
q_\ell \wedge q_m \rightarrow {\cal O}(1), 
\qquad
q_\ell \circ q_m \sim {\cal O}(\epsilon^{-2}) \rightarrow \infty.
\label{limit2}
\eea
In this limit, the modified Bessel function ${\bf K}_n$ exhibits 
the following asymptotic behavior:
\bea
 {\bf K}_n \Big(m z \Big) \rightarrow \sqrt{ {\pi \over 2 m z}} e^{-m \vert
z \vert} 
\left[ 1+ {\cal O} \l({1 \over m \vert z \vert}\r) \right].
\nonumber
\eea
Most remarkably, the asymptotic behavior is {\sl independent} of 
the index $n$.  Compared to the planar effective action, there is not
the extra $n!$ factor that (partially) cancels $\rm 
N!$ in the denominator,
which shows that the summed form of the effective action markedly changes.
In the low-energy limit, the Fourier-transformed 
kernels, $\widetilde{{\cal K}_n}$, obey the following recursive relation:
\bea
\widetilde{{\cal K}^{}}_{n+1} \left( { k} \circ { k} \right) = 
\left( \vert\theta \cdot {k} \vert \over m \right)
\, \widetilde{{\cal K}^{}}_n \left( {k} \circ {k} \right),
\nonumber
\eea
viz.
\bea
\widetilde{{\cal K}_n} \left({k} \circ {k} \right)
= \left( { \vert \theta \cdot {k} \vert \over m} \right)^n 
\, \widetilde{{\cal Q}^{}} \left({k} \circ {k} \right).
\nonumber
\eea
Here, the kernel $\widetilde{{\cal Q}_{}}$ is given by:
\bea
\widetilde{{\cal Q}_{}} \left({k} \circ {k} \right)
= \left( 2 \pi \right)^{(1 - d)/2} \left\vert {1 \over m \, 
\theta \cdot {k}} \right\vert^{1 \over 2}  \exp \Big(- m \vert 
\theta \cdot {k}  \vert \Big).
\nonumber
\eea
Note that, in the power-series expansion of the effective action, a natural 
expansion parameter is $\vert \theta \cdot {k} \vert$ .

Thus, the nonplanar one-loop effective action in momentum space 
is expressible as:
\bea
 \Gamma_{\rm np}[\Phi] &=&{\hbar \over 2} \int {\d^d k \over (2\pi)^d} 
\widetilde{\cal K}_{-{d \over 2}} \left({k} \circ {k} \right)
\sum_{{\rm N}=2}^{\infty} \sum_{n=1}^{\rm N-1}
\left(- {\lambda \over 4 m }\right)^{\rm N}
\nonumber \\
&\times & \left( {1 \over n!} \vert \theta \cdot {k} \vert^n 
\left[ \widetilde{\Phi} \cdots \widetilde{\Phi} \right]_{\star_n}[k] \right)
\left( {1 \over ({\rm N} - n)!} \vert\theta \cdot {k}\vert^{{\rm N} - n} 
\left[ \widetilde{\Phi} \cdots \widetilde{\Phi} \right]_{\star_{{\rm N} - n}}
[-k] \right).
\label{seriesform}
\eea
Utilizing the relation between the generalized $\star_n$ products and the
{\sl scalar} open Wilson line operators, as elaborated in \cite{ours},
the nonplanar one-loop effective action can be summed up into a 
remarkably simple closed form. Denote the rescaled coupling parameter as
$g := \lambda / 4m$ (see Eq.(\ref{scalarwilson})). Then, 
because of the algebraic relation 
$[\widetilde{\Phi} \star_0 \widetilde{\Phi}]_{k} = (2 \pi)^d \delta^{(d)} (k)$,
domain of the double summations can be extended to $n=0$, N$=0$ terms, as 
they yield identically vanishing contribution {\sl after} $k$-integration is
performed. Once this arrangement is made, partial summations over N and $n$ 
can be performed explicitly. Exploiting the exchange symmetry 
$n \leftrightarrow ({\rm N} - n)$, the summation domain $(n, $N) over the 
lower triangular lattice points can be mapped to the one over the upper 
triangular lattice points.  By averaging over the two summation domains, one 
can then rearrange the double summations into {\sl decoupled} ones over $n$ 
and $({\rm N}-n)$.  One finally obtains: 
\bea
\Gamma_{\rm np}[\Phi] = 
    {\hbar \over 2} 
\int {\d^d k \over (2 \pi)^d}\,
 W_{k}[\Phi] \cdot \widetilde{{\cal K}_{-{d \over 2}}}
\left({ k} \circ {k} \right) \cdot W_{-{k}}[\Phi],
\nonumber
\eea
yielding precisely the aforementioned result, Eq.(\ref{npresult}).

\section{Conclusions and Discussions}\label{sec5}
\setcounter{section}{5}
\setcounter{equation}{0}
\indent 

In this paper, we have studied the one-loop effective action in the
noncommutative $\lambda[\Phi^3]_\star$-theory. In order to make direct 
comparison with the Seiberg-Witten limit of the open string worldsheet formulation, 
in computing the one-particle irreducible one-loop Green functions, 
we have adopted the worldline 
formulation of the theory. We have observed that, at low-energy, the 
one-loop diagrams, both planar and nonplanar, are completely factorizable.
We have shown explicitly that, while the planar contribution is expressed in
terms of the standard $\star$-products, the nonplanar contribution 
is expressible solely in terms of the generalized $\star$-products. This 
implies that structure of the one-loop effective action reveals quite 
a different physics between
planar and nonplanar contributions. In particular, we were able to show that,
at large noncommutativity limit, the nonplanar contribution is expressible
in terms of {\sl open} Wilson line operators, thus completing the proof of
our earlier result in \cite{ours}. The planar contribution, on the other 
hand, gives rise to a noncommutative version of the Coleman-Weinberg type 
potential, in contrast to the anticipation that the planar part ought to
be expressible in terms of Wilson loop opeators -- 
the putative master field in planar limit of matrix-valued field theories. 
The next obvious step is to extend the computation to two-loop and confirm 
that two-loop effective action is re-expressible in terms of (at most) three
open Wilson line operators. We will report the result in a separate
publication.   

Our computation in the worldline formulation has revealed several new light
concerning spacetime interpretation of the one-loop physics.
Among them is concerning the shift of the momentum integration variable, as is 
evident from integration $K$ in section 2.2. It suggests that variation of 
the
internal momemtum $\Delta k^a$ integrated along the entire loop amounts to 
\bea\label{vari}
\Delta\int\limits_0^\T \d \tau \, k^a(\tau) &=& {i\over2}\int\limits_0^\T 
\d \tau \Bigl\{\, {\dot x}^a(\tau) -{1 \over 2} \sum_{j=1}^{\rm N}
\theta^{ab} p_{b \, j} \nu_j\delta(\tau-\tau_j)\,  \Bigr\}\nn\\
&=& {i\over2}\sum_{j=1}^{\rm N} \theta^{ab}p_{b \, j} \eps_j \ .
\eea
Recalling $\eps_j$ take either 0 or 1, depending on whether the $j$-th vertex
is untwisted or twisted insertion, we recognize that  
the above relation is precisely the momentum-space counterpart of the
dipole relation Eq.(\ref{relation}). Recalling that $k^a (\tau)$ is the 
conjugate momentum to $x^a(\tau)$, both of which are associated with the 
virtual quanta circulating around the loop, the above relation asserts 
that the virtual quanta is not a point-like constituent, obeying the standard 
Fourier transformation relation between $x^a(\tau)$ and $k^a(\tau)$, but 
behaves as a sort of rigid rod whose size is proportioanl to the momentum.
We trust details of this unusual physics --- physics of dipoles --- 
deserve further investigation and intend to report new understanding concerning 
this aspect in separate publications.  

\section*{Acknowledgement}
We thank H. Liu, P. Mayr, Y. Oz, and L. Susskind for discussions.  
SJR acknowledges warm hospitality of Henri Poincar\'e Institut, Institut
des Hautes Etudes Scientifiques, Theory Division at CERN, and Department
of Mathematics and Physics at Amsterdam University at various stage of 
this project.
 
\appendix
\section*{Appendix}
\section{Schwinger parametrization of one-loop Feynman diagram}
\setcounter{equation}{0}
\indent 

In this section, we will provide a check point of the worldline formulation 
introduced in sections 2 with the standard Feynman diagrammatics. 
Although a general expression for N-point, one-loop Feynman diagrams are
given, for instance, in \cite{Roiban} or in 
\cite{seiberg1}, they are not in a convenient form for comparison with 
the results in the worldline formulation, mainly because of different 
moduli parametrizations and omission of overall normalization and 
combinatoric factors (which are necessary for resummation of the N-point
Green functions into the effective action). 
Let us consider the N-point Feynman diagram (cf. Fig. 2), wherein 
the nonplanar phase-factors 
$e^{ik\wedge p_1}$, $e^{i(k+p_1)\wedge p_2}$, $e^{i(k+p_1+p_2)\wedge p_3}$, 
$\cdots$, $e^{ik\wedge p_{\rm N}}$ are inserted at each of the N interaction 
vertices, respectively, as well as the overall Filk's phase-factor 
$e^{-{i\over2}\sum_{i<j}p_i\wedge p_j}$. The one-loop Feynman amplitude
is given by 
\bea\label{FN}
{}~F_{\rm N} =e^{-{i\over2}\sum_{i<j}p_i\wedge p_j}
\int{\d^d k\over(2\pi)^d} {e^{ik\wedge P}
e^{i\sum_{l=2}^{\rm N-1}\sum_{i=2}^{l-1} p_i\wedge p_l \eps_l }
\over k^2 (k+p_1)^2 (k+p_1+p_2)^2 \cdots (k+ p_1 + p_2 + \cdots
+ p_{\rm N-1})^2 }. 
\eea
Here, the twist factor $\eps_i=1$ or 0; $i=1,\cdots,\rm N$ 
are inserted for the planar the and nonplanar vertex insertions,
respectively, and
\bea\label{bigP}
P^a =\sum_{i=1}^{\rm N} \eps_i \, p^a_i \ .
\eea
We rewrite the momentum integral in terms of an overall modulus integral 
(global Schwinger parameter $T\equiv\tau_{\rm N}$) and $\rm (N-1)$ relative 
moduli integrals (local Schwinger parameters, $\tau_i$): 
\bea\label{FN2} 
{}~F_{\rm N}&=&\prod_{i < j = 1}^{\rm N} 
e^{-{i\over2} p_i\wedge p_j}
\prod_{k = 2}^{\rm N-1} \prod_{\ell =2}^{k-1} e^{i p_\ell \wedge p_k \eps_k}
\\
&\times& \int\limits_0^\infty \d \T \left({1\over4\pi \T}\right)^{d\over2}
\prod_{n=1}^{\rm N-1}\int\limits_0^{\tau_{n+1}}\d \tau_n 
\exp\left[\,{1\over \T}\left(\sum_{j=1}^{\rm N-1}\tau_{j\, j+1}
\sum_{\ell=1}^j p_\ell +{i\over2}\theta \cdot P\,\right)^2
- \sum_{j=1}^{\rm N-1}\tau_{j\, j+1}\Bigl(\sum_{\ell=1}^j p_\ell \Bigr)^2\,
\right]\ ,
\nonumber
\eea
where $\tau_{ij} = \tau_i - \tau_j$.
From the energy-momentum conservation, the following relations can be deduced:
\bea
P\circ P = -\sum_{i<j}^{\rm N} \eps_{ij}^2\, p_i\circ p_j \ ,
\nonumber
\eea
\bea
\sum_{j=1}^{\rm N-1}\tau_{j \, j+1}\sum_{l=1}^j p_\ell \wedge P 
=
\sum_{i<j}^{\rm N}(\tau_i+\tau_j)\eps_{ji}p_i\wedge p_j
=
-\sum_{i<j}^{\rm N} \sum_{k=1,k\not=i,j}^{\rm N} 
\tau_k\, \eps_{ji}p_i\wedge p_j \ ,
\nonumber
\eea
\bea
{1\over \T}\left(\sum_{j=1}^{\rm N-1}\tau_{j \, j+1}\sum_{\ell=1}^jp_\ell\,
\right)^2
- \sum_{j=1}^{\rm N-1}\tau_{j\, j+1}\Bigl(\sum_{\ell=1}^jp_\ell \Bigr)^2 
=\sum_{i<j}^{\rm N} p_i \cdot {\cal G}_B(\tau_j,\tau_i) \cdot p_j\ ,
\nonumber
\eea
and
\bea\label{ide4}
\sum_{i<j}^{\rm N} \eps_{ij}p_i\wedge p_j=
\sum_{i<j}^{\rm N} (\nu_i-\nu_j)p_i\wedge p_j=
\sum_{i<j}^{\rm N} \tau_{ij}p_i\wedge p_j=0 \ .
\eea
Utilizing the identity Eq.(\ref{ide4}), the nonplanar phase-factor can be simplified as
\bea
e^{i\sum_{l=2}^{\rm N-1}\sum_{i=2}^{l-1}p_i\wedge p_l\eps_l} 
=e^{i\sum_{i<j}\eps_i p_i\wedge p_j}\ ,
\nonumber
\eea
thereby deriving
\bea\label{FN3}
{}~F_{\rm N} = e^{-{i\over2}\sum_{i<j}\nu_j \, p_i\wedge p_j}
\int\limits_0^\infty \d \T \left({1\over4\pi \T}\right)^{d\over2}
\prod_{n=1}^{\rm N-1}\int\limits_0^{\tau_{n+1}} \d\tau_n 
\exp\left[\,\sum_{i<j}^{\rm N} p_i \cdot {\cal G}_{B\theta}(\tau_i,\tau_j;
\eps_i,\eps_j) \cdot p_j \,\right]\ .
\eea
Using the above identities again, Filk's overall phase-factor 
in Eq.\eq{FN3} can be reduced to
\bea
-{i\over2}\sum_{i<j}p_i\wedge p_j\nu_j = 
-{i\over2}\sum_{i<j}p_i\wedge p_j{1 \over 2} (\nu_i+\nu_j)\ ,
\nonumber
\eea 
and 
\bea\label{FN4}
{}~F_{\rm N} &=& 
\int\limits_0^\infty \d \T \left({1\over4\pi \T}\right)^{d\over2}e^{-m^2\T}
\prod_{n=1}^{\rm N-1}\int\limits_0^{\tau_{n+1}} \d \tau_n \,
\prod\limits_{i<j}^{\rm N} \exp \left( {i\over2} p_i\wedge p_j\,{1 \over 2} 
(\nu_i+\nu_j)\eps(\tau_{ji}) \right)
\nn\\
&&\times\exp\Bigl[\,\sum_{i<j}^{\rm N} p_i \cdot 
{\cal G}_{B\theta}(\tau_i,\tau_j;\eps_i,\eps_j) \cdot p_j \,\Bigr]\ .
\eea
Here, having in mind of generalization to all possible
orderings, we have attached 
$\eps(\tau_{ji})$ to the phase-factor in order to blindly count the 
ordering dependences. If we shuffle the external legs, we obtain different 
diagrams with different phase-factors. Thus all ordered integrals correspond 
to all possible distinct Feynman diagrams. In this way, we recover the full 
integral regions after summing all contributions for fixed $\nu_i$. 
Obviously, we then also have to sum over all combinations of $\nu_i$. 
As, in section 2.2,  Eq.\eq{GNfinal} is invariant under the simultaneous 
constant shifts 
of all $\tau$ (by noticing Eq.\eq{GBtheta} and Eq.\eq{ide4}), we can fix 
one of them in Eq.\eq{GNfinal}; for example, $\tau_{\rm N} = \T$, 
as has been assumed throughout this section, and reproduce 
\bea 
\Gamma_{\rm N} &=& C (-\lambda)^{\rm N} \sum_{\nu_j}\,
\int \d \T e^{-m^2 \T} \left({1\over4\pi \T}\right)^{d\over2}
\left(\prod_{i=1}^{\rm N-1}\int\limits_0^\T \d \tau_i\right)
e^{{i\over4}\sum_{i<j}p_i\wedge p_j (\nu_i+\nu_j)\eps(\tau_{ij}) } \nn\\
&&\times
\exp\Bigl[\,{1\over2}\sum_{i,j=1}^{\rm N} p_i \cdot 
{\cal G}_{B\theta} (\tau_i,\tau_j;\eps_i,\eps_j)\cdot p_j\,\,\Bigr]\ .
\nonumber
\eea
Here, $C$ is the normallization factor defined by 
a fraction between the symmetry factor $S_{\rm N}$ and the number of 
topologically distinct integration regions $C_{\rm N}$:
\bea\label{combi}  
C= {S_{\rm N} \over C_{\rm N}}\ .
\eea
Following closely the arguments of Ref~\cite{exact}, we will now determine
the combinatoric factor $C_{\rm N}$. 
If one expands the effective action in powers
of the coupling constant, the combinatoric weight for the expansion of the 
Feynman diagrams is obtained by shuffling the external interaction vertices.
One obtains
\bea\label{w}
w={S_{\rm N} \, n^\T \over \rm N!}\ ,
\eea
where $n^\T$ is the number of the topologically distinct diagrams out of those 
shuffled diagrams, $S$ is the symmetry factor for those. In the present 
case, this number $w$ is given by  
\bea\label{wvalue}
w={1\over 2 \cdot \rm N \cdot 2^N} \ .
\eea
It comprises of trace-log factor ${1\over2}$, 
the ${1\over \rm N}$ coming from the Taylor expansion of an one-loop 
form $\ln(1+x)$, and the coupling factor $2^{\rm -N}$. 
(One may simply multiply $2^{\rm -N}$ to the result Eq.(3.25) 
of Ref~\cite{exact}). Combining Eq.\eq{w} and Eq.\eq{wvalue}, we obtain 
\bea
S_{\rm N} = { (\rm N-1)!\over 2^{\rm N+1}\, n^\T}\ .
\nonumber
\eea 
As $C_{\rm N}$ is defined by 
\bea
C_{\rm N}={({\rm N} -1)! \over n^\T}\ ,
\nonumber
\eea
we can put these into Eq.\eq{combi} and obtain
\bea
C={1\over 2\cdot \rm 2^N}\ ,
\nonumber
\eea
therefore proving that our world-line master formula coincides with 
the Feynman diagrammatics result. 

The commutative limit, $\theta^{ab}\ra0$, also comes out correct, as 
all the $\rm 2^N$ terms of the $\left\{\nu_i \right\}$ summation are 
reduced to the same, single contribution, reproducing the commutative
result.


\end{document}